\documentclass[preprint]{revtex4}
\usepackage[utf8]{inputenc}

\usepackage{siunitx}
\usepackage{graphicx}
\usepackage{amssymb}

\begin{document}

\title{Scattering contrast in GHz frequency ultrasound subsurface atomic force microscopy for detection of deeply buried features}

\date{May 2020}

\author{Maarten H. van Es}
\email[email:]{maarten.vanes@tno.nl}
\affiliation{Optomechatronics, TNO, Delft, The Netherlands}
\author{Benoit A.J. Quesson}
\affiliation{Acoustics and Sonar, TNO, The Hague, The Netherlands}
\author{Abbas Mohtashami}
\author{Daniele Piras}
\author{Kodai Hatakeyama}
\affiliation{Optomechatronics, TNO, Delft, The Netherlands}
\author{Laurent Fillinger}
\author{Paul L.M.J. van Neer}
\affiliation{Acoustics and Sonar, TNO, The Hague, The Netherlands}



\begin{abstract}
While Atomic Force Microscopy is mostly used to investigate surface properties, people have almost since its invention sought to apply its high resolution capability to image also structures buried within samples. One of the earliest techniques for this was based on using ultrasound excitations to visualize local differences in effective tip-sample stiffness caused by the presence of buried structures with different visco-elasticity from their surroundings. While the use of ultrasound has often triggered discussions on the contribution of diffraction or scattering of acoustic waves in visualizing buried structures, no conclusive papers on this topic have been published. Here we demonstrate and discuss how such acoustical effects can be unambiguously recognized and can be used with Atomic Force Microscopy to visualize deeply buried structures.
\end{abstract}

\maketitle

\section{Introduction}
\label{sec:intro}

Subsurface Atomic Force Microscopy (AFM) --- a set of techniques which enables to visualize buried nano-scale features using AFM --- has garnered a lot of interest \cite{Hu2011ImagingSubsurfaceStructures,Killgore2011QuantitativeSubsurfaceContact,Kolosov1996NanoscaleImagingMechanical,Es2018MappingBuriedNanostructures,Cretin1993ScanningMicrodeformationMicroscopy,Cantrell2007NanoscaleSubsurfaceImaging}. In fact, imaging buried structures with the nanometer scale resolution of scanning probe microscopy has a wide range of applications running from metrology in the semiconductor industry \cite{Es2018SoundingOutBuried,Stan2016NanoscaleTomographicReconstruction}, to investigating processes in live cells \cite{Tetard2008ImagingNanoparticlesCells}. While there has been quite some debate over the contrast mechanism(s) which enable subsurface imaging \cite{Es2018MappingBuriedNanostructures,Cantrell2007NanoscaleSubsurfaceImaging,Verbiest2017SubsurfaceContrastDue}, elasticity (or more properly, visco-elasticity) is generally cited as the main physical cause of the contrast \cite{Killgore2011QuantitativeSubsurfaceContact,Es2018SoundingOutBuried,Rabe1994AcousticMicroscopyAtomic,Parlak2008ContactStiffnessFinite,Rabe2012AtomicForceAcoustic}. Such elasticity-based subsurface AFM has shown few nanometer resolution on buried structures \cite{Es2019QuantitativeTomography}, but the depth sensitivity of this method is limited by the amount of stress that can be applied without damaging the sample. However, many applications require the ability to image more deeply buried features than is feasible with elasticity based subsurface AFM. In semiconductor metrology for example, 3D NAND memory is quickly growing in the number of layers of memory cells, necessitating metrology measurements through stacks which are already many micrometers thick. Elasticity based subsurface AFM cannot detect small structures buried so deeply in the relatively stiff materials used in semiconductor manufacturing. However, a different contrast mechanism for performing subsurface AFM has been suggested in one paper by Hu et al. \cite{Hu2011ImagingSubsurfaceStructures} based on scattering of acoustic waves by buried structures. Because in this case the information about the structures is carried to the surface by the waves, this mechanism promises much better results for deeply buried structures. However, there has never been a follow up to this publication to our knowledge and in fact, as we will describe below, one would expect to observe a number of distinctive effects due to the scattering process which have not been described by Hu et al. Here we will present high signal-to-noise ratio (SNR) images that provide conclusive evidence that indeed scattering of acoustic waves at over \SI{1}{GHz} frequencies can be used as contrast mechanism, by highlighting different diffraction effects we observed. We do so using structures buried several micrometer below the surface to emphasize the ability to detect deeply buried structures. 

As recognized by Hu et al.\cite{Hu2011ImagingSubsurfaceStructures}, the acoustic wavelength employed is key to enable contrast based on scattering of acoustic waves. For particles which are significantly smaller than the wavelength, the scattered energy is given by the theory of Rayleigh and scales with $\frac{d^6}{\lambda^4}$\cite{cobbold2006foundations}, with $d$ the feature size and $\lambda$ the acoustic wavelength. As a result, for wavelengths longer than the feature size, the contribution from scattering quickly becomes insignificant compared to the background, unperturbed acoustic wave. A non-linear detection scheme using frequency mixing \cite{Verbiest2015BeatingBeatsMixing}, as typically used in subsurface AFM, only exacerbates this problem. Therefore, to employ scattering as contrast mechanism, the wavelength should at most be of similar size as the features of interest. At \SI{1}{GHz} the acoustic wavelength in silicon is approximately \SI{10}{\micro\meter}, a length scale which is easily accessible to AFM. Therefore, like Hu et al. \cite{Hu2011ImagingSubsurfaceStructures} we have chosen to focus on an acoustic frequency of about \SI{1}{GHz} to show scattering as contrast mechanism in subsurface AFM. As discussed in the methods and discussion sections, if scattering of acoustic waves indeed has a significant contribution to the measurement results, we expect to see a number of unique effects from that, that have never been reported yet with subsurface AFM. For example, from the theory of acoustics, we expect complex changes in the diffraction pattern with frequency \cite{Faran1951} and diffraction ripples outside the patterned area \cite{cobbold2006foundations}.  Here we will investigate in detail through theory and simulations what effects to expect and experimentally verify whether these effects take place. Thus we will determine whether scattering of acoustic waves is the dominant contrast mechanism in our experiments and how to unequivocally recognize this. 

In this paper, we will first discuss again briefly the experimental setup (see ref.~\cite{Neer2019OptimizationAcousticCoupling} for a detailed description) enabling repeatable, acoustic scattering based subsurface AFM in section~\ref{section:results}. Details on the experimental implementation can be found in section~\ref{sec:methods}. Next, we will explain the types of experiments performed in section~\ref{section:results:exps} and then the simulations we implemented, together with the basics of acoustics that are needed to understand the observed effects in section~\ref{section:results:sims}. Finally, we will show and discuss the various, distinctive effects that confirm the occurrence of scattering in sections~\ref{section:results:coupling}, \ref{section:results:scans} and~\ref{sec:discussion}. All together, these results demonstrate the potential of acoustic scattering based subsurface AFM to visualize deeply buried structures non-destructively.

\section{Results}
\label{section:results}
\begin{figure}
	\centering
	\includegraphics[width=0.8\linewidth]{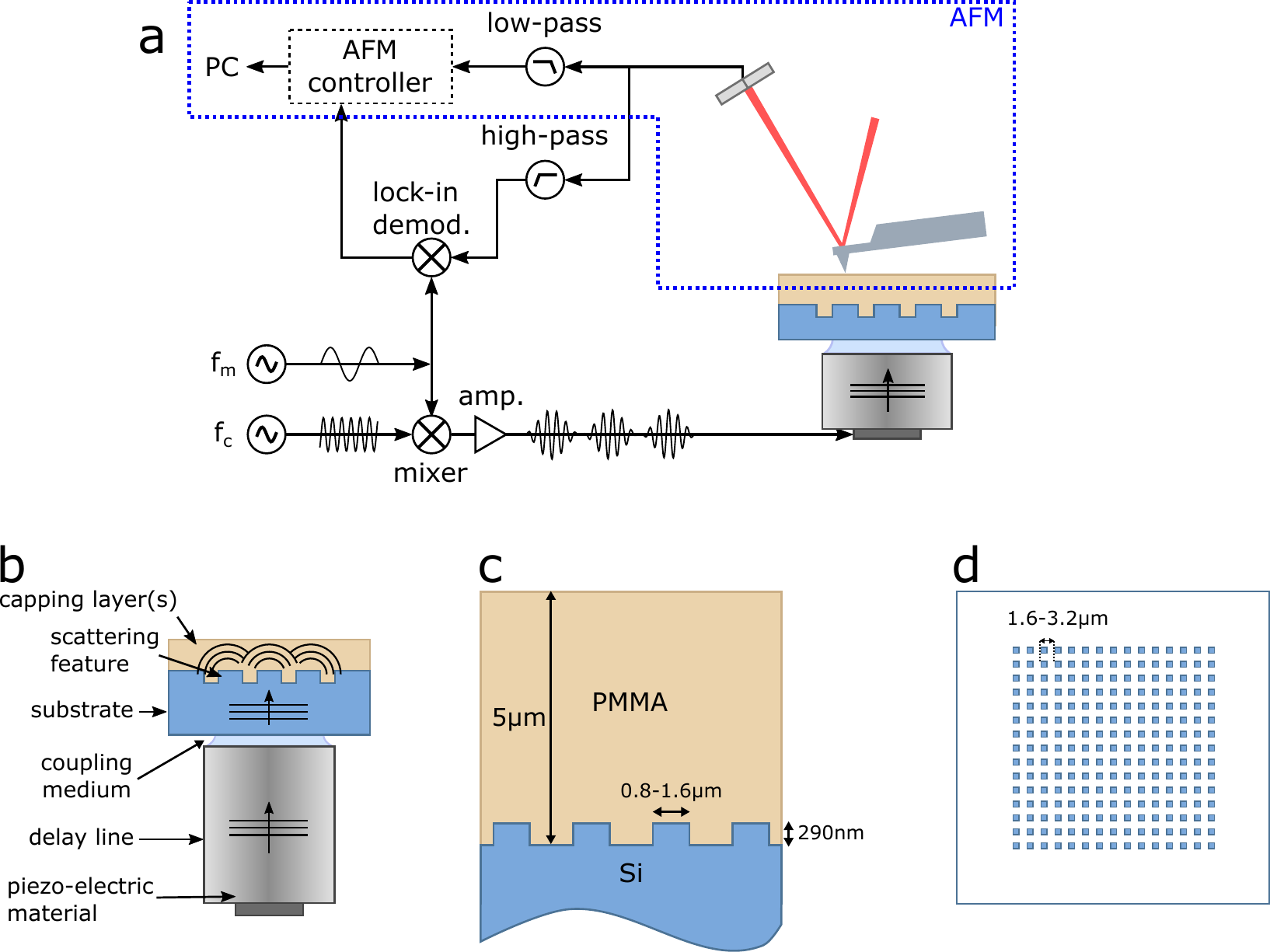}
	\caption{Schematic showing setup and sample. (a) Concept of the experimental setup, showing: signal generation through mixing the carrier and modulation frequencies $f_c$ and $f_m$; power amplification through an amplifier, \textit{amp.}; sound generation in the piezoelectric transducer; sound propagation through the delay line, the coupling fluid and sample with buried features; the AFM tip and AFM system for sensing the sound; and demodulation with a lock-in amplifier using $f_m$ as reference signal. (b) a more detailed view of the acoustic stack showing plane acoustic waves scattering off the buried features in the sample. (c) and (d) cross-section and planar lay-out of the top of the sample showing the configuration of the buried features.}
	\label{fig:setupandsample}
\end{figure}

The measurement setup and sample are sketched in Fig.~\ref{fig:setupandsample}. As in many visco-elasticity based subsurface AFM setups, a piezoelectric ultrasound transducer is placed under the sample. The resulting out-of-plane oscillations of the sample surface are picked up using an AFM probe \cite{Rabe1994AcousticMicroscopyAtomic,Binnig1986AtomicForceMicroscope}. A water layer is used for efficiently coupling the ultrasound energy from the transducer to the sample. Using this setup, various measurements have been performed and compared to simulations. First of all, simulations and measurements of sweeping the excitation frequency have been performed to, respectively, predict and characterize the acoustical behavior of the piezoelectric transducer-coupling layer-sample stack, hereafter referred to as acoustic stack or simply stack. This allows to identify standing wave resonances inside the acoustic stack and thereby to optimize the experiment for contrast and SNR. Secondly, imaging measurements and simulations have been performed that highlight the various effects that occur due to scattering.

\subsection{Overview of experiments}
\label{section:results:exps}
Three types of measurement were performed sequentially: (1) coupling layer thickness estimation using ultrasound pulse-echo analysis, (2) characterization of the standing waves in the complete acoustic stack using carrier frequency sweeps, to be compared with the 1D Krimholtz, Leedom and Matthae (KLM) model (see section~\ref{section:results:sims}) and (3) AFM scans, which can be compared to the 2D finite element method (FEM) simulations (see section~\ref{section:results:sims}). These three types of measurements are further described in more details below.

\subsubsection{Coupling layer thickness estimation}
A liquid coupling layer is necessary between the transducer and sample to ensure efficient acoustical energy transfer. We chose water because of its relatively good acoustical properties, low evaporation rate and non-toxicity. Still, the attenuation of high frequency sound in water leads to a requirement of sub micrometer thickness for this layer\cite{Neer2019OptimizationAcousticCoupling,Quesson2018}, and the formation of standing waves in the acoustic stack leads to a requirement for the temporal stability of the layer thickness of better than \SI{10}{nm} during measurements \cite{Neer2019OptimizationAcousticCoupling,Quesson2018}. These requirements were tackled by the design of a custom sample clamp. We also monitored the thickness and stability of the coupling layer thickness prior to an AFM measurement as described in detail in \cite{Neer2019OptimizationAcousticCoupling,Quesson2018}. In short, we monitored the resonance frequency due to the coupling layer in a pulse-echo measurement scheme. A \SI{50}{ns} long, \SI{1}{GHz} bandwidth linear frequency modulated pulse centered at \SI{1.25}{GHz} was used for that purpose. The monitoring of the coupling layer thickness was done every \SI{10}{s} and went on until a stabilization was observed with a coupling layer thickness below \SI{1}{\micro\metre} and variability under \SI{10}{nm}. 

\subsubsection{Carrier frequency sweep}
A continuous amplitude-modulated (AM) signal using a carrier frequency in the GHz order and a modulation frequency equal to the cantilever's first contact resonance mode was used. The carrier frequency was swept in order to identify the local stack resonances that are expected to occur due to the multiple interfaces in the acoustic stack. Since it was expected from the simulations that the contribution of the coupling layer would lead to MHz order resonance spacing (see Fig.~\ref{fig:stackresonances}, the carrier frequency sweeps were done using a \SI{1}{MHz} step. The carrier frequency sweep was performed while measuring with the AFM at a single position next to the area with features to avoid damaging the sample surface at the feature location. Once the frequency corresponding to the maximum cantilever amplitude was selected, an AFM scan was performed above the features.

\subsubsection{AFM Scans}
In order to measure the out-of-plane displacements on top of the sample, the AFM was operated in contact mode. Since the probe was used to measure the surface vibrations and not to probe its stiffness or elasticity, the contact force was set relatively low at around \SI{10}{nN} (compared to other subsurface measurement methodologies). On the one hand, such force level enhances the tip-sample non-linearity and the `effective Q' factor of the cantilever, enhancing SNR, and on the other hand it minimizes the influence of contributions from elastic deformation to the observed contrast. In addition, it minimizes the mechanical damage to the sample surface due to the contact between tip and sample. The amplitude and the phase of the down-mixed signal are recorded along with the topography information (height and deflection error) through the auxiliary inputs on the AFM controller. In order to minimize artefacts from scanning, the sample is always rotated at around \ang{45} with regards to the scan axes such that the scanning direction does not align with feature directions in the sample.

\subsection{Overview of simulations}
\label{section:results:sims} 
In order to understand the interaction between the ultrasound waves and the sample under investigation, two different types of simulation have been performed. First, a 1D KLM \cite{Leedom1971,Krimholtz1972} model is used to evaluate an approximate frequency response of the acoustic stack. This response is dominated by resonances due to standing waves within the stack. Knowing this response enables to select optimal frequencies during the experiment. Then a 2D Finite Element Model (FEM) is used to evaluate the additional effect of the scattering  from finite size features at selected frequencies to describe the experimentally observed wavefront patterns on the top surface of the sample. In both models the silicon delay line and the sample are in ideal contact (no water layer); therefore the silicon delay line and the silicon substrate behave as a single, thicker, layer. For simulations including the water layer, we refer to our earlier paper \cite{Neer2019OptimizationAcousticCoupling} dealing in detail with this aspect of the system.

In the 1D KLM model, all the non-piezoelectric layers in the sample stack are definedby their thickness, density and speed of sound. The piezoelectric transducer is additionally characterized by its piezoelectric coefficient and relative permitivity. All the layers are modelled as transmission lines, and the signal is collected at surface of the most distant layer from the transducer.

\begin{figure}
	\centering
	\includegraphics[width=0.5\linewidth]{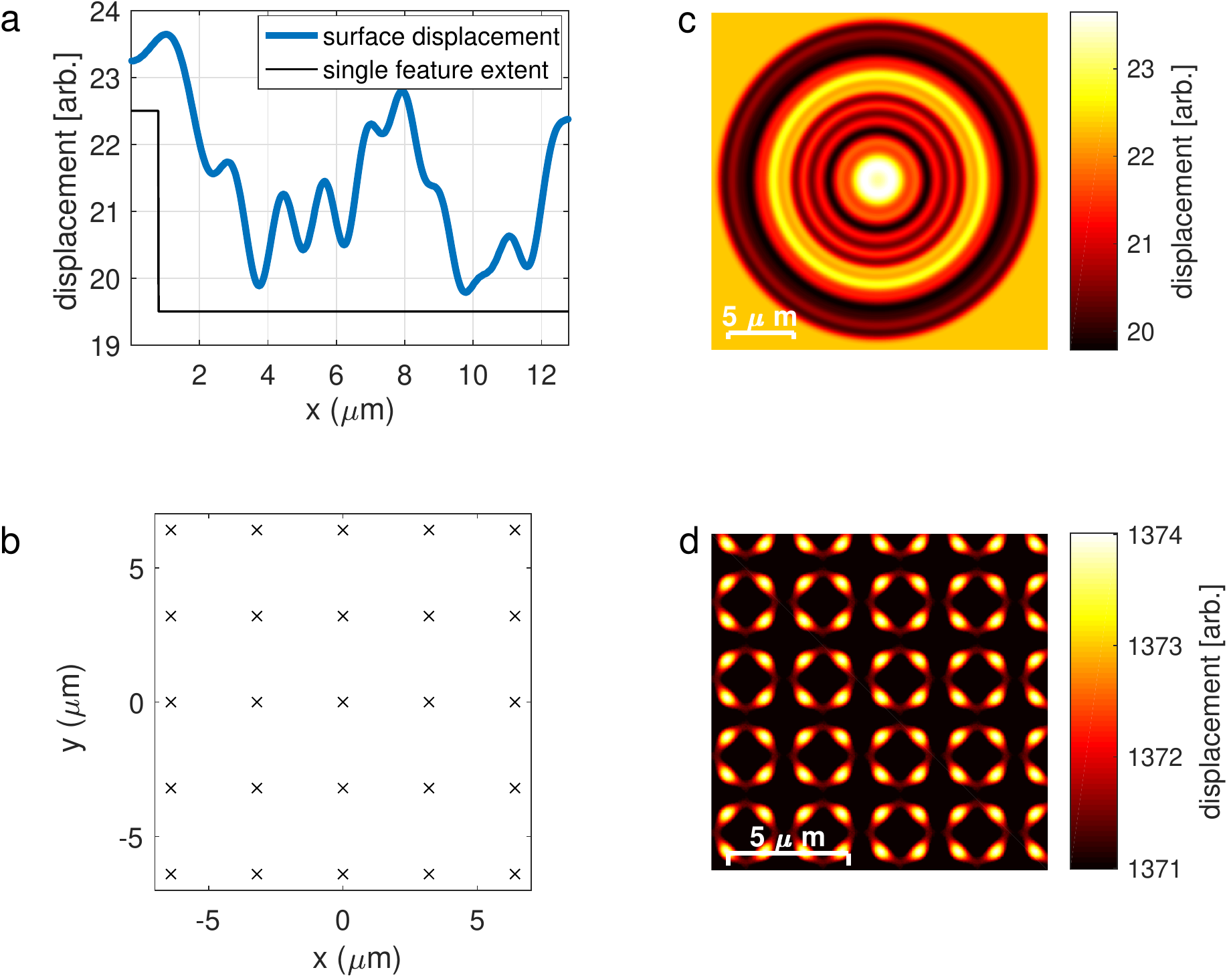}
	\caption{Surface displacement pattern simulation steps. (a) Top surface vertical (z) displacement (thick blue line) obtained by 2D simulation of the scattering by a single feature (extent shown by the thin black line). (b) Simulated imaging point spread function obtained by application of a cylindrical symmetry to the simulated surface displacement. (c) Feature positions. (d) Surface displacement pattern resulting from the scattering by the feature array, obtained by convolution of the point spread function (b) with the array (c).}
	\label{fig:sim2d}
\end{figure}

The 2D FEM analysis allows to visualize the effect of any inhomogeneities in the sample stack on the standing  wave pattern. In particular, the features forming the interface between the Silicon substrate and the PMMA capping affect the wave propagation in the sample. If scattering occurs at these sites, this scattered ultrasound field also contributes to the standing wave pattern, so that the acoustic energy interferes in the vertical direction between the layers and in the horizontal direction between the features of the grating. To efficiently evaluate the surface displacement pattern (and thus to select an appropriate experimental carrier frequency) the FEM analysis in COMSOL is first performed with a 1D array of 15 features to capture the feature-to-feature contribution on top of the standing wave pattern. For computational constraints, the analysis is performed in 2D (so that effectively the features are infinitely long trenches rather than square pits). A custom perfectly matched layer at the side boundaries of the medium is implemented in order to limit the analysis to a smaller size sample (approximately 1.3 mm wide) compared to an actual die for AFM measurements (5 to 10 mm wide). In this way it is possible to evaluate at which frequency to expect the maximum contrast in displacement between on and off feature, as well as the spatial distribution of the displacement field. However, because they involve trench-like features, these 2D simulations cannot be used to reproduce the experimental surface displacement resulting from pillar-like features with finite size in all three directions. The 3D character of the experiment needs to be accounted for. To overcome numerical limitations of the FEM model, this 3D simulation is approximated using a 2D simulation performed with a single feature, extrapolating the result to a  point spread function (PSF) by applying a rotational symmetry, and finally simulating the response to a feature array by convolution of the PSF with the feature array pattern. This is illustrated in Fig.~\ref{fig:sim2d}. Such convolutions are presented for comparison with the experimentally obtained AFM scans in Fig.~\ref{fig:scanspoffertjeswaffels}c and d.

\subsection{Coupling of the transducer with the sample}
\label{section:results:coupling}

\begin{figure}
	\centering
	\includegraphics[width=0.5\linewidth]{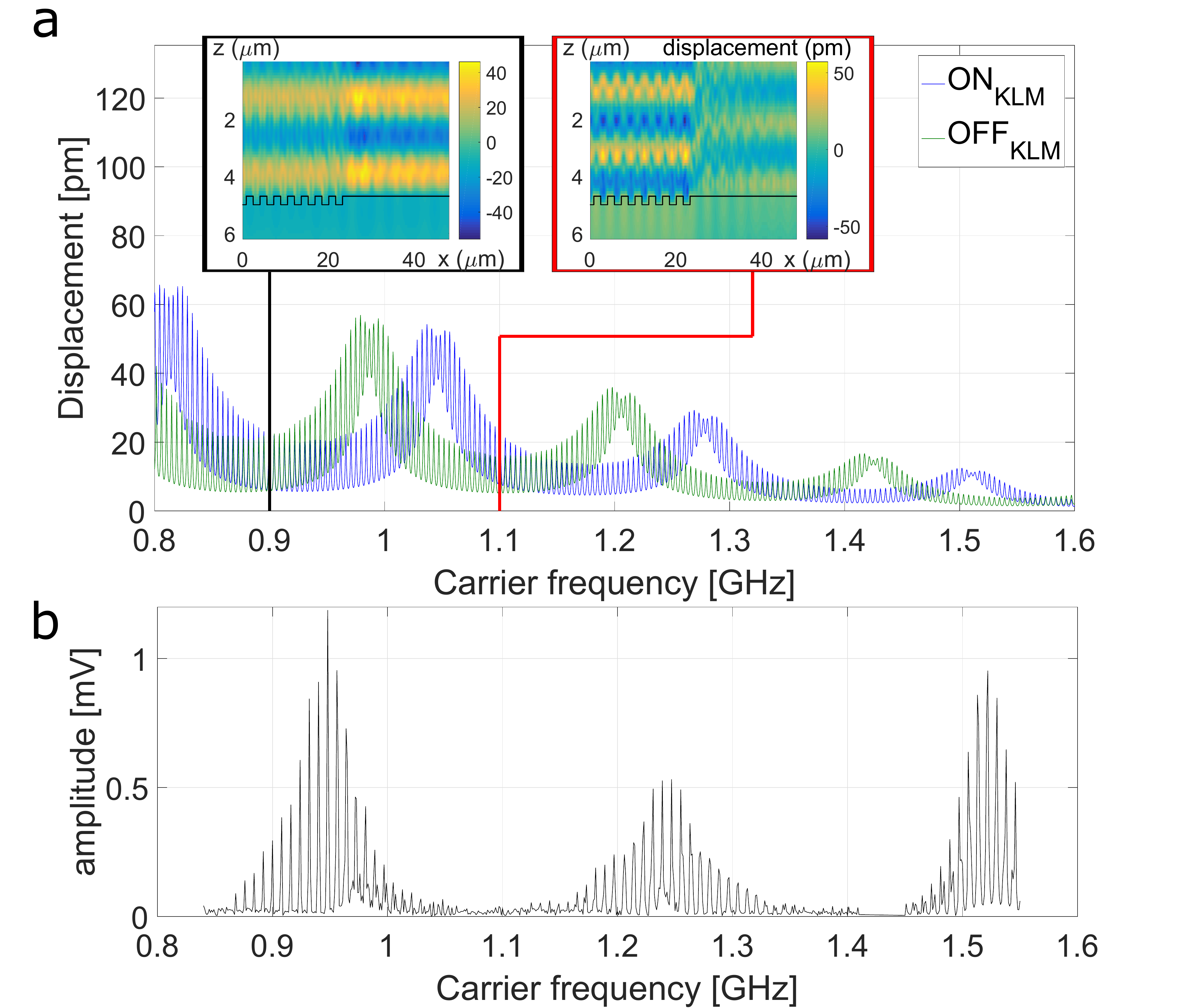}
	\caption{
	Frequency spectra --- (a) from simulations and (b) from measurement --- revealing standing wave resonances in the acoustic stack. (a) Shows the stack resonances as measured from the top surface displacement amplitude in the simulations on top of the buried structures (blue) or next to the buried structures (green). Each of the spectra shows two periodicities --- a small one (about \SI{4}{Mhz}) from standing waves in the silicon delay line and a large one (about \SI{230}{MHz}) from standing waves in the top PMMA layer. Due to a different thickness of PMMA on top of and next to the buried features, the large periodicity differs. The insets show a cross-section of the simulated displacement field within the PMMA at two different frequencies, highlighting the different resonance conditions between the different locations. (b) Shows the measured cantilever amplitude spectrum (after frequency mixing) on a sample patch with PMMA and without buried structures. It exhibits a periodicity of about \SI{230}{MHz}, again from the PMMA, and of about \SI{8}{MHz}, from the individual delay line and sample. Because of the water coupling layer in between, these cannot be considered as fused together in the actual experiment as opposed to the simulations, and therefore the standing wave conditions happens for smaller wavelengths, thus resulting in larger frequency spacings.
	}
	\label{fig:stackresonances}
\end{figure}

Fig.~\ref{fig:stackresonances}a shows a wide frequency sweep performed with the KLM model on top of and next to the features (blue and green respectively). This spectrum exhibits resonances from standing waves that present two orders of periodicity. The large periodicity (with a period around \SI{230}{MHz}) is due to the PMMA top layer. This is expected when an odd integer number of quarter wavelength fit in the PMMA layer thickness --- that is, they should occur every $f_{\frac{3\lambda}{4}}-f_{\frac{1\lambda}{4}}=\frac{c_{p,PMMA}}{2h_{PMMA}}=\SI{232}{MHz}$. The small periodicity (with a period around \SI{4}{MHz}) results from standing waves in the silicon part of the sample stack, which are expected when an integer number of half wavelengths fit in the silicon layer thickness. In this simulation, the sample and delay-line are considered together as one piece of silicon, and we expect the half wavelength resonance in the silicon layers to be $f_{Si,\lambda/2}=\frac{c_{p,Si}}{2h_{Si}}=\SI{4.5}{MHz}$. In the experiment the water layer splits the silicon layer almost in half, so we actually expect a half wave resonance more around \SI{8}{MHz}, as confirmed in the experimental frequency sweeps presented in Fig.~\ref{fig:stackresonances}b. Figure \ref{fig:stackresonances}b also confirms the \SI{230}{MHz} periodicity due to the PMMA layer in the experiment. The experimental data are composed of 10 smaller carrier frequency sweeps acquired separately over the course of two consecutive days (highlighting the stability and reproducibility of the setup and experiments). The three resonance peaks at \SI{0.948}{GHz}, \SI{1.247}{GHz} and \SI{1.522}{GHz} match well with the expected resonances for the PMMA layer. More detailed graphs and analyses of these resonance effects can be found in our earlier paper \cite{Neer2019OptimizationAcousticCoupling}.

The insets in Fig.~\ref{fig:stackresonances}a show a vertical cross section of the simulated displacement field over and next to the feature array at two different frequencies. These insets show how the contrast and displacement patterns depend in complex ways on the chosen carrier frequency.

\subsection{AFM imaging scans}
\label{section:results:scans}
\begin{figure}
	\centering
	\includegraphics[width=0.5\linewidth]{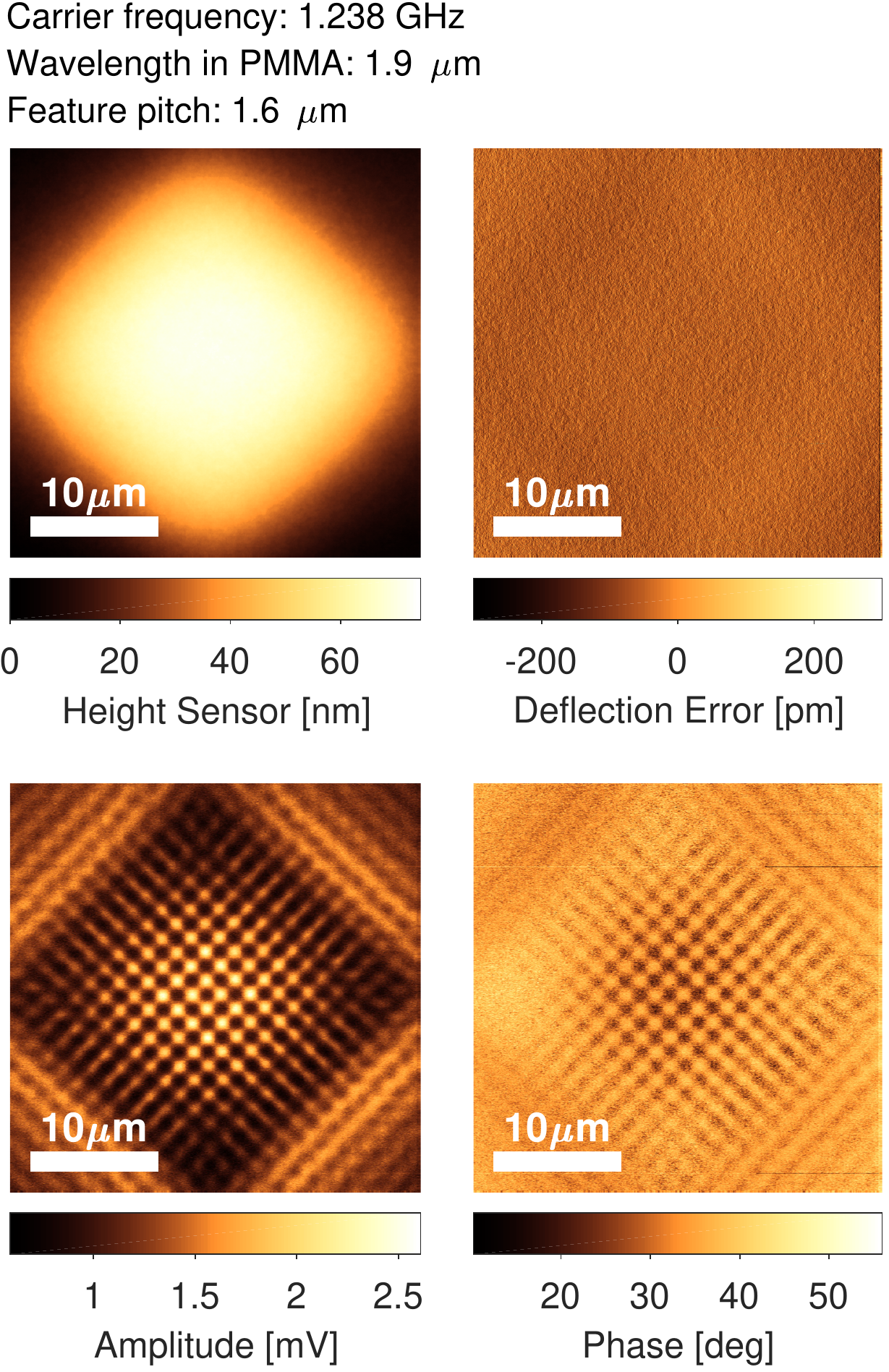}
	\caption{Topography (top left), feedback error (top right), cantilever amplitude (bottom left) and cantilever phase (bottom right) recorded over an array of features with pitch \SI{1600}{nm}. Carrier frequency was \SI{1.238}{GHz}, corresponding to a wavelength in PMMA of approx. \SI{1900}{nm}. The topography shows an overall print-through of the underlying matrix of features, but the individual features cannot be discerned in this image. In contrast, the phase and especially the amplitude show intricate diffraction patterns that extend even beyond the physical extent of the feature array.
	}
	\label{fig:scansedge}
\end{figure}

Figure \ref{fig:scansedge} shows the result of an AFM scan over \SI{800}{nm} subsurface features (\SI{1600}{nm} pitch) at a carrier frequency of \SI{1.238}{GHz} (wavelength in PMMA: approx. \SI{1900}{nm}). The two top plots show the topography and deflection error and the bottom plots show the corresponding down-mixed amplitude and phase of the acoustic standing wave pattern at the surface. The scan area is \SI{32x32}{\micro\meter} with 1024 measurement points per line and a scan rate of \SI{0.5}{Hz}. Since the sample surface is not planarized, an overall print-through up to about \SI{70}{nm} is observed in the topography due to the underlying subsurface array. However, we note that the individual features are not visible and smoothed out in the topography. In contrast, the amplitude and the phase channels clearly show distinctive features linked to the array of the subsurface features below. Outside of the apparent location of the array of features, ripples are visible in the amplitude image. Especially near the edge of the array, also additional fine structure can be discerned in the amplitude image. These observations suggest that the observed pattern has contributions from a large area of the buried feature pattern and not only from the part directly below the tip.

\begin{figure}
	\centering
	\includegraphics[width=0.5\linewidth]{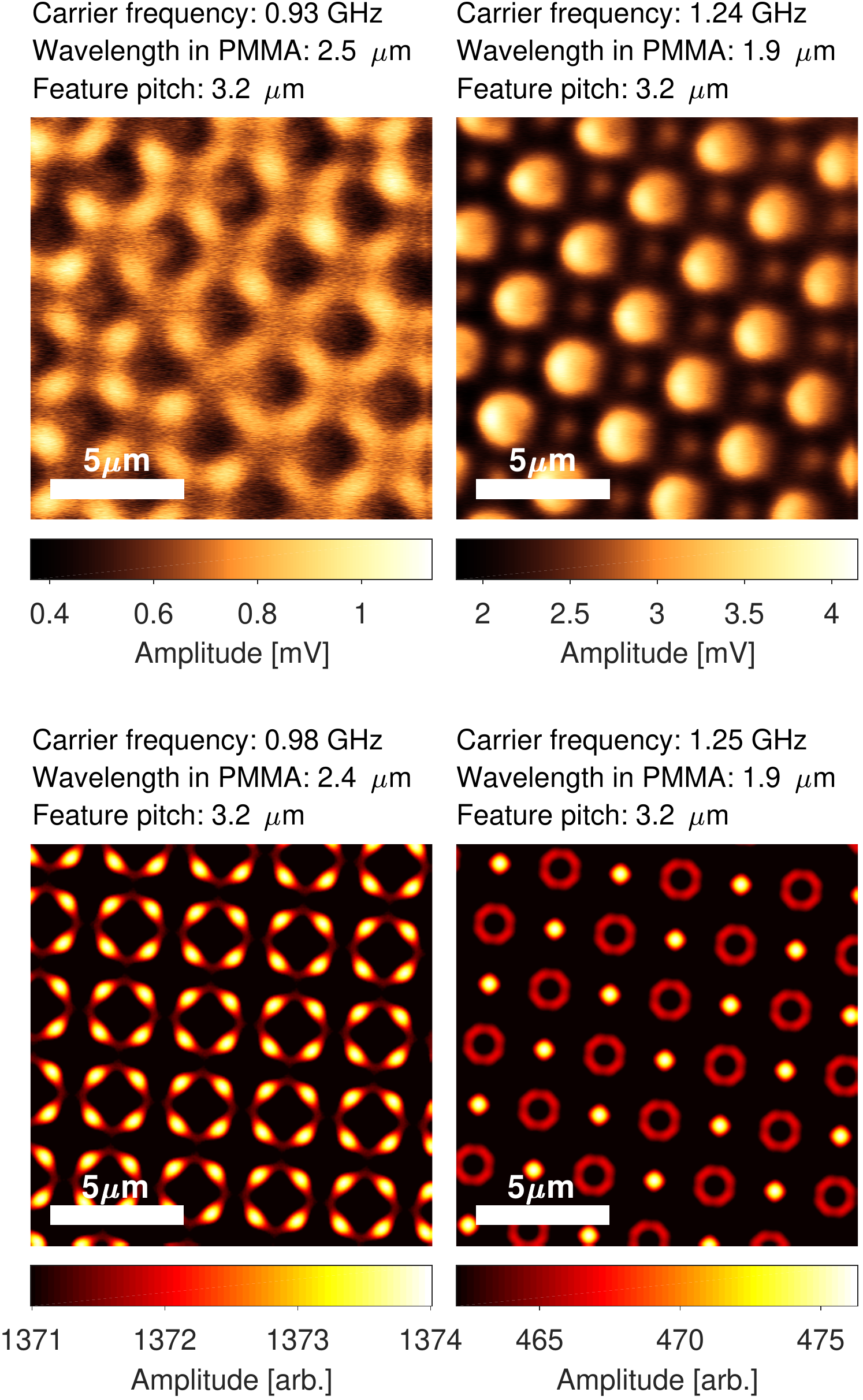}
	\caption{Measured (top) and simulated (bottom) amplitude images of an array of features with pitch \SI{3200}{nm}. The measured images were recorded back to back at the same location but at different carrier frequencies: the left images were made using lower frequencies than the right images. Similar patterns may occur at slightly different frequencies in the simulations and the experiments as the simulations do not include the coupling water layer and assume a certain stack geometry and material properties.
	}
	\label{fig:scanspoffertjeswaffels}
\end{figure}

Fig.~\ref{fig:scanspoffertjeswaffels}a and b show the results of two scans over \SI{1600}{nm} features (pitch \SI{3200}{nm}) at two different carrier frequencies of \SI{1.238}{GHz} and \SI{0.9325}{GHz}, respectively. These two measurements were performed back to back at the same location. In order to resolve the interference patterns better, the overall scan size was reduced to \SI{15x15}{\micro\meter} with the remaining scan settings unchanged from the previous, larger area scan. One sees a distinct pattern change between the two different carrier frequencies. Simulation results from the 2D FEM analysis show patterns that are remarkably similar (Fig.~\ref{fig:scanspoffertjeswaffels}c and d), when keeping in mind the simplifications done in the simulation. Note that, except from choosing a carrier frequency to match the patterns, no fitting was performed. The carrier frequencies are expected to be slightly different between experiments and simulations because the simulations do not include the water layer between transducer and sample and, of course, material properties and the experimental geometry are not known exactly.

\section{Discussion}
\label{sec:discussion}
The following observations, which are discussed further below, demonstrate that scattering is the dominant contrast mechanism for the experiments described in this paper: 
\begin{enumerate}
    \item The depth of the buried features and the good match between scattering based simulations and the experiments;
    \item The presence of ripples in the measurements outside of the extent of the matrix of features (Fig.~\ref{fig:scansedge});
    \item The dependence of sample surface displacement patterns on the carrier frequency;
    \item The dependence of sample surface displacement amplitude on the carrier frequency and on the acoustic stack geometry and material properties.
\end{enumerate}

1. Depth of the features ---
As discussed in the introduction, the dominant contrast mechanism identified in the literature is viscoelastic contrast. The measurable depth range depends on the extent of the stress field resulting from contact with the AFM tip. As a rule of thumb the depth range is about 3 times the contact radius, according to Hertzian contact mechanics applied to the tip - sample contact \cite{Hu2011ImagingSubsurfaceStructures}. The cantilever used in this work (\SI{0.4}{N/m}, ScanAsyst-Air, Bruker) has a specified nominal tip radius of \SI{2}{nm} and a specified maximum tip radius of \SI{12}{nm}. The applied static force was \SI{10}{nN}. Assuming Hertzian contact mechanics, the contact radius is calculated to be \SI{1.8}{nm} and \SI{3.2}{nm} for the nominal and maximum radius, respectively. Using the aforementioned rule of thumb, the measurable depth range for viscoelasticity based contrast is less than \SI{10}{nm}. The features measured in this work were buried below a \SI{5}{\micro\meter} PMMA layer. Therefore, the measured subsurface contrast cannot be caused by viscoelasticity based contrast: the sample features are buried too deep. In contrast, the measurements shown in Fig.~\ref{fig:scansedge} show a clear representation of the features in the subsurface amplitude and phase channels. In addition, the good match between the measurements and the simulations (see Fig.~\ref{fig:scanspoffertjeswaffels}) supports acoustic scattering as the source of contrast.

2. Ripples caused by diffraction ---
The measurements of the matrix of square features (see Fig.~\ref{fig:scansedge}) show a ripple like effect outside of the area where the matrix of features is located. This phenomenon is typical for diffraction and hence an indicator that the contrast mechanism is caused by scattering effects.
It can be qualitatively described according to acoustic theory as follows: The incident compressional wave coming from below the features is scattered or diffracted and travels further upwards to the sample surface. This scattered wavefront may be described as originating from a set of virtual sources at the feature locations, each with the same amplitude and phase because the incident wave is approximately planar over the size of the feature matrix. Now the diffraction pattern of a mono-frequency sound source can be calculated at each spatial location by taking the Fourier transform of the aperture, and multiplying the result by the amplitude and phase of the source excitation. Therefore, taking the Fourier transform of the rectangular envelope of the matrix of features, we obtain a sinc-function as expected envelope of the diffraction ripples along each axis of the matrix of features. This sinc-like nature of the diffraction pattern is clearly visible in Fig.~\ref{fig:scansedge}. Furthermore, because the \SI{1.9}{\micro\meter} acoustic wavelength is only slightly larger than the \SI{1.6}{\micro\meter} feature pitch, and because the \SI{24}{\micro\meter} extension of the feature matrix and the \SI{5}{\micro\meter} depth imply the measurement location is in the near-field
, it can be expected that representations of the individual features are visible in the diffraction pattern inside and also slightly outside the rectangular envelope of the matrix of features.
Note that we used here the definition for near-field as commonly used in acoustics or antenna design: it means roughly that part of the field that is in front of the natural focus of a source\cite{cobbold2006foundations}. This is very different from the definition that is perhaps most well-known in the field of AFM (as used for example in `Scanning Near-Field Optical Microscopy'), where it means roughly that part of the field that is so close to the source that it is necessary to consider non-radiating solutions to the field equations --- this is more equivalent to what is called the `reactive near field' in antenna design.

3. Dependence on the carrier frequency ---
Another key indicator for the scattering based contrast mechanism is the fact that the shape of the measured representation of the matrix of features varies with the carrier frequency (see Fig.~\ref{fig:scanspoffertjeswaffels}). This is caused by a change in the diffraction pattern as a function of frequency. These effects could also be modeled accurately by the wave propagation simulations, as is indicated by the good match between the measurement data and the wave propagation simulations (see Fig.~\ref{fig:scanspoffertjeswaffels}). This frequency dependency could furthermore not be explained by a visco-elastic contrast mechanism.


4. Dependence on the acoustic stack geometry and material properties ---
The measured and simulated carrier frequency spectra show peaks and valleys with frequency periods in the order of approximately \SI{8}{MHz} (see Fig.~\ref{fig:stackresonances}). The peaks and valleys correspond to successive thickness resonances caused by the interference of acoustic waves propagating inside the acoustic stack (the piezomaterial, the delay line, sample and the coupling layer). For a more detailed explanation the reader is referred to \cite{Neer2019OptimizationAcousticCoupling}. The aforementioned frequency dependent interference pattern of the out-of-plane sample surface displacement is a clear indication that it is caused by acoustic effects, rather than by stiffness effects.

In the measurement and simulation results of Fig.~\ref{fig:stackresonances} a frequency periodicity of approximately \SI{230}{MHz} can be observed. This periodicity corresponds to the quarter wavelength acoustic standing wave behavior in the PMMA top layer (given a compressional wave speed of \SI{2324}{m/s} in the PMMA and a \SI{5}{\micro\meter} PMMA thickness we expect a \SI{232.4}{MHz} period). 

In conclusion, we have shown that scattering of acoustic waves is the dominant contrast mechanism in the subsurface AFM measurements presented here. The scattering based character of the contrast is supported by a number of observations based on simulations and measurements. This includes the observation of diffraction ripples outside the region directly above the scatterers and significant changes in the observed diffraction pattern upon changing the ultrasound carrier frequency. The observations are accompanied by matching results from modelling without needing fitting parameters. These results point to some of the challenges in measured data interpretation - namely that the observed pattern cannot be directly interpreted in terms of the buried pattern - but foremost demonstrate the potential of this technique to visualize deeply buried structures non-destructively using AFM.

\section*{Acknowledgements}

We would like to thank Walter Arnold for helpful discussions on the concept and implementation of the experiments described here.

This article received funding from the TNO early research program 3D nano-manufacturing.

\section{Methods}
\label{sec:methods} 

The setup consisted of a waveform generator (M8195A, Keysight, Santa Rosa, USA) where custom excitation signals were defined. The electrical signal was then amplified with a power amplifier (ZHL-2-8+, Mini-circuits, New-York, USA) and divided with a power splitter (ZFRSC-42S+, Mini-circuits, New-York, USA). One signal path was routed to an oscilloscope (DSA 70804B, Tektronix, Beaverton, USA) for monitoring the excitation signal. The other signal path was routed through a circulator (PE8400, Pasternack, Irvine, USA) and a piezoelectric transducer (custom design, Kibero, Saarbrücken, Germany) where the electric signal was converted to an acoustic wave. The circulator allowed to monitor the echo's coming back to the piezo on the oscilloscope also. The acoustic wave traveled from the piezo through a \SI{450}{\micro\metre} silicon delay line, a sub-micrometer coupling water layer  and a custom sample (Kavli Nanolab, Delft, NL). The piezoelectric transducer, the coupling layer and the sample were maintained in relative position by a custom clamp \cite{Neer2019OptimizationAcousticCoupling,Quesson2018}. The sample consisted of a \SI{525}{\micro\metre} thick silicon substrate and a \SI{5}{\micro\metre} thick PMMA layer. Before adding the PMMA, feature arrays were etched in the silicon containing $15\times15$ square wells with a duty cycle of \SI{50}{\percent}, a depth of \SI{290}{nm} and pitches of \SIrange{1600}{3200}{nm}.

The out-of-plane oscillations were picked up on top of the PMMA layer using an AFM probe (ScanAsyst-Air, 0.4 N/m, Bruker, Billerica, USA) in combination with an AFM system (Bruker Dimension Icon, Bruker, Billerica, USA). Due to the non-linear tip-sample interaction, frequency mixing occurs \cite{Verbiest2015BeatingBeatsMixing} and the cantilever is excited at frequencies corresponding to linear combinations of the frequencies supplied to the ultrasonic transducer. In particular, the cantilever is excited at the modulation or difference frequency. Choosing the modulation frequency to be equal to the contact resonance is advised in order to amplify the measured signal by the cantilever's Q factor. The cantilever response is detected using the optical beam deflection system of the AFM in combination with an external lock-in amplifier (UHFLI, Zurich Instruments, Z\"urich, Switzerland). Finally, the resulting amplitude and phase signals are transferred to a computer to be recorded along with the corresponding topography information from the AFM.

\section{Data Availability}
The data that support the findings of this study are available from the corresponding author upon reasonable request. 

\section{Author Contributions}

All authors were involved in discussions on the concept, implementation and analysis of the experiments and simulations described in this paper, as well as in manuscript preparation. More specifically, M.H.E. performed early experiments (not published here), and oversaw the manuscript preparation. B.A.J.Q., A.M. and K.H. performed the measurements for which the results are published here. D.P. and L.F. implemented the simulations. P.L.M.J.N. developed the theory and the concept for an improved experimental setup.


\end{document}